\documentclass[a4paper,twoside]{article}
\pdfoutput=1
\usepackage{epsfig}
\usepackage{subfigure}
\usepackage{calc}
\usepackage{amssymb}
\usepackage{amstext}
\usepackage{amsmath}
\usepackage{amsthm}
\usepackage{multicol}
\usepackage{pslatex}
\usepackage{apalike}
\usepackage{SCITEPRESS}
\usepackage[small]{caption}
\usepackage{setspace}
\usepackage{epstopdf}
\usepackage{pbox}
\usepackage{array}
\usepackage{fancyhdr}

\fancypagestyle{firststyle}
{
   \fancyhf{}
   
   \fancyhead[L]{\textbf{11th International Conference on Informatics in Control, Automation and Robotics\\
   Vienna University of Technology\\
   September 1-3, 2014, Vienna, Austria\\
   http://www.icinco.org/}}
}

\begin{document}

\thispagestyle{firststyle}

\title{Vehicle Parameter Independent Gain Matrix Selection for a Quadrotor using State-Space Controller Design Methods}

\author{\authorname{Graeme N. Wilson, Alejandro Ramirez-Serrano, and Qiao Sun}
\affiliation{Department of Mechanical and Manufacturing Engineering, University of Calgary, 2500 University Drive NW, Calgary, Alberta, T2N 1N4}
\email{\{gnw.wilson\}@gmail.com, \{aramirez, qsun\}@ucalgary.ca}
}


\abstract{With quadrotor use seeing extensive growth in recent years, the autonomous control of these Unmanned Aerial Vehicles (UAVs) is an increasing relevant and intersting field. In this paper a linear state-space approach at designing a stable hover controller in the presence of disturbances is presented along with simulation of control system performance. Additionally the design of a tracking system, for linear inertial position and yaw, is presented with simulation results. The gain matrix developed for this control system is independent of the specific quadrotor parameters, meaning that this same gain matrix can be used on a wide variety of quadrotors without modification. The hover and tracking controllers designed in this paper proved to perform well in simulation under perturbation disturbances and normally distributed disturbances on the UAVs linear speeds and angular speeds.} 

\onecolumn \maketitle \normalsize \vfill

\section{\uppercase{Introduction}}
\label{sec:introduction}

\noindent The use of UAVs (Unmanned Aerial Vehicles) has seen extensive growth in recent years for industrial, military, and consumer use. One of the most common types of UAVs, although a naturally unstable system, is the quadrotor. Quadrotors are typically operated by remote control using a joystick \cite{Fernando2013}. There has also been an increasing interest in autonomous control of a quadrotor as seen in research in recent years.

Autonomous PID control of a quadrotor has been tested by \cite{Jun2013} and \cite{Harandi2010} where each orientation angle (pitch, roll, yaw) had an individual PID controller that was experimentally tuned. These two control approaches, as well as another controller by \cite{Zhan2012}, dealt with PID control only for attitude stabilization. Another PID controller by \cite{Li2011} added position stabilization as well as attitude stabilization. In the work by \cite{Bai2012} robust PID control was developed to deal with disturbances, although it only dealt with attitude stabilization. In a different approach to PID tuning, pole selection was performed using transfer function analysis to produce an attitude and position stabilized controller \cite{Sa2013}.

Using state space-methods for quadrotor control, the work by \cite{Reyes-Valeria2013} produced two gain matrices for a gain scheduled controller using LQR gain selection. This work used one gain matrix when the quadrotor was far away from the trajectory, and a second matrix when the quadrotor was on the desired trajectory; this work did not deal with disturbances \cite{Reyes-Valeria2013}. In a comparison of control methods \cite{Al-Younes2010} tested a PID controller, LQR controller, and non-linear Adaptive Integral Backstepping Controller for attitude stabilization; this work did not deal with trajectory tracking.

This paper presents a linear state-space method of control system design for the purpose of attitude and position stabilization as well as path tracking. The gain matrix that is designed within this work is independent of the vehicle's properties, which mean it is valid for any quadrotor configuration (provided it holds to the standard general design of a quadrotor, four fixed rotors with a mass in center).

\section{\uppercase{Modeling}}
\begin{figure}[h]
\centering
\includegraphics[width=2.5in]{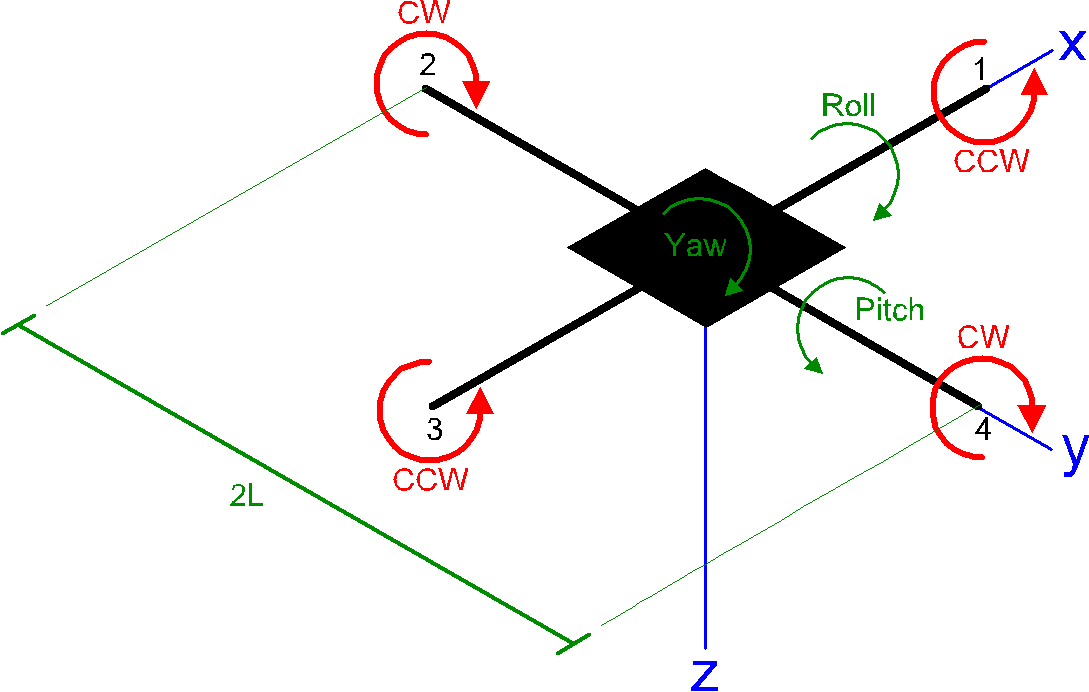}
\caption{Quadrotor Diagram.}
\label{fig:QuadrotorDiagram}
\end{figure}

\noindent To model the system the quadcopter configuration shown in Figure \ref{fig:QuadrotorDiagram} was used. In Figure \ref{fig:QuadrotorDiagram} it can be seen that roll is counterclockwise about the \textit{x-axis}, pitch is counterclockwise about the \textit{y-axis}, and yaw is counterclockwise about the \textit{z-axis}. Additionally Rotors 1 and 3 rotate counterclockwise, while Rotors 2 and 4 rotate clockwise.

\subsection{Non-Linear Dynamics}
\noindent To derive the dynamic equations for this quadrotor first the rotation matrix from Inertial (North East Down at initial position) to Body frame (Shown on Figure \ref{fig:QuadrotorDiagram}) is defined (Equation (\ref{eq:RotationMatrix})). The roll \(\phi\) is the rotation counterclockwise about the x-axis, the pitch \(\theta\) is the rotation counterclockwise about the y-axis, and the yaw \(\psi\) is the rotation counterclockwise about the z-axis.

\begin{equation}
\footnotesize
\label{eq:RotationMatrix}
\centering
R_{IB}
=
\left[
\begin{array}{ccc}
c\theta c\psi & c\theta s\psi & -s\theta \\
s\phi s\theta c\psi - c\phi s\psi & s\phi s\theta s\psi + c\phi c\psi & s\phi c\theta \\
c\phi s\theta c\psi + s\phi s\psi & c\phi s\theta s\psi-s\phi c\psi & c\phi c\theta

\end{array}
\right]
\end{equation}

Next the set of generalized coordinates is set to be \((x,y,z,\omega_x,\omega_y,\omega_z)\), where \(x,y,z\) are the position component from initial position measured in the body axis, and \(\omega_x,\omega_y,\omega_z\) are the angular velocity components in the body axis. Lagrangian modeling techniques were then used to produce the non-linear dynamic equations of the system shown in Equation (\ref{eq:non-linearModel}) where the values for \((w_1, w_2, w_3, w_4)\) are defined in Equation (\ref{eq:wDefintion}). These equations use the quadrotor parameter \(k\) for rotor speed torque constant, \(b\) for rotor speed to drag constant, \(m\) for quadrotor mass, \(L\) for the half distance between the rotors, \(g\) is the gravitational constant, and \((I_x,I_y,I_z)\) are the components of the inertia matrix (principle axes as shown in Figure \ref{fig:QuadrotorDiagram}). The values \(\omega_{M_1},\omega_{M_2},\omega_{M_3},\omega_{M_4}\) (in Equation (\ref{eq:wDefintion})) are the four rotors angular speeds respectively. The values \(F_{D_x},F_{D_y},F_{D_z}\) represent the disturbance forces to the systems that also include aerodynamic effects lumped in with the disturbances. The values \(\tau_{D_x}, \tau_{D_y}, \tau_{D_z}\) are the disturbance torques to the system which include the aerodynamics torques and gyroscopic torques lumped in with the disturbances. This was done to simplify the modeling while still considering all internal and external disturbances. The linearized state space model does not consider the disturbance terms. In this study the rotor inertia's are neglected (assumed to be small).

\begin{equation}
\label{eq:non-linearModel}
\centering
\left[
\begin{array}{c}
\ddot{x} \\ 
\ddot{y} \\ 
\ddot{z} \\ 
\dot{\omega_x} \\ 
\dot{\omega_y} \\ 
\dot{\omega_z}
\end{array}
\right]
=
\left[
\begin{array}{c}
-gs{\theta} + F_{D_x}/m \\
gs{\phi}c{\theta} + F_{D_y}/m \\
gc{\phi}c{\theta} + w_1 \left(k/m\right) + F_{D_z}/m \\
-w_2 \left(kL/I_x\right) + \tau_{D_x}/I_x \\
-w_3 \left(kL/I_y\right) + \tau_{D_y}/I_y \\
w_4 \left(bL/I_z\right) + \tau_{D_z}/I_z
\end{array}
\right]
\end{equation}

\begin{equation}
\label{eq:wDefintion}
\left[
\begin{array}{c}
w_1 \\
w_2 \\
w_3 \\
w_4 
\end{array}
\right]
=
\left[
\begin{array}{c}
\omega_{M_1}^2 + \omega_{M_2}^2 + \omega_{M_3}^2 + \omega_{M_4}^2 \\
\omega_{M_2}^2 - \omega_{M_4}^2 \\
\omega_{M_1}^2-\omega_{M_3}^2 \\
\omega_{M_1}^2 - \omega_{M_2}^2 + \omega_{M_3}^2 - \omega_{M_4}^2
\end{array}
\right]
\end{equation}

While these non-linear equations represent the system well, it is desirable to linearize them to simplify the control system design. 

\subsection{Linearized Dynamics}
\noindent To linearize Equation (\ref{eq:non-linearModel}) first the traditional small angle approximations are made (\(\sin{x} \approx x, \, \cos{x} \approx 1\)) followed by substitutions for \(u_1,u_2,u_3\) and \(u_4\) from Equation (\ref{eq:uDefintion}).

\begin{equation}
\label{eq:uDefintion}
\left[
\begin{array}{c}
u_1 \\
u_2 \\
u_3 \\
u_4 
\end{array}
\right]
=
\left[
\begin{array}{c}
g+w_1\left(k/m\right) \\
-w_2\left(kL/I_x\right) \\
-w_3\left(kL/I_y\right) \\
w_4 \left(bL/I_z\right)
\end{array}
\right]
\end{equation}

These substitutions satisfy linearization of Equation (\ref{eq:non-linearModel}); however, it is desirable to include the states for linear speed \((\dot{x},\dot{y},\dot{z})\) as well as Euler angular rates \((\dot{\phi},\dot{\theta},\dot{\psi})\) since we would like to control these as well. Considering the Euler angular rate relationships in Equation (\ref{eq:eulerRates}), it is clear these need to be linearized as well.

\begin{equation}
\label{eq:eulerRates}
\left[
\begin{array}{c}
\dot{\phi} \\
\dot{\theta} \\
\dot{\psi} 
\end{array}
\right]
=
\left[
\begin{array}{c}
\omega_x+\left(\omega_y\sin{\phi}+\omega_z\cos{\phi}\right)\\
\omega_y\cos{\phi}-\omega_z\sin{\phi}\\
\left(\omega_y\sin{\phi}+\omega_z\cos{\phi}\right) / \cos{\theta})
\end{array}
\right]
\end{equation}

Once again using small angle approximation and using i) the product of two small angles is near zero, ii) the product of a small angle and an angular rate is small, a rough linearization for the Euler angular rates can be defined as \(\dot{\phi} \approx \omega_x\), \(\dot{\theta}\approx\omega_y\), and \(\dot{\psi}\approx\omega_z\). This final linearization leads to the linearized state-space model for the quadrotor in the form \(\dot{X}=AX+BU\) defined in Equation (\ref{eq:linearStateSpaceModel}). Note that to save space the \(AX\) and \(BU\) terms have been multiplied together to form vectors (Equation (\ref{eq:linearStateSpaceModel})). Furthermore, the \(A\) and \(B\) matrices can be recovered by knowing that the state vector \(X\) is \((\dot{x},\dot{y},\dot{z},x,y,z,\omega_x,\omega_y,\omega_z,\phi,\theta,\psi)^T\), and the input vector \(U\) is \((u_1,u_2,u_3,u_4)^T\). Note that this input vector \(U\) is used to solve for the rotor speeds (\(\omega_{M_1},\omega_{M_2},\omega_{M_3},\omega_{M_4}\)) when controlling the quadrotor.

\begin{equation}
\label{eq:linearStateSpaceModel}
\left[
\begin{array}{c}
\ddot{x} \\
\ddot{y} \\
\ddot{z} \\
\dot{x} \\
\dot{y} \\
\dot{z} \\
\dot{\omega}_x \\
\dot{\omega}_y \\
\dot{\omega}_z \\
\dot{\phi} \\
\dot{\theta} \\
\dot{\psi} 
\end{array}
\right]
=
\left[
\begin{array}{c}
-g\theta \\
g\phi \\
0\\
\dot{x} \\
\dot{y} \\
\dot{z} \\
0\\
0\\
0\\
\omega_x \\
\omega_y \\
\omega_z \\
\end{array}
\right]
+
\left[
\begin{array}{c}
0 \\
0 \\
u_1 \\
0 \\
0 \\
0 \\
u_2 \\
u_3 \\
u_4 \\
0 \\
0 \\
0 \\

\end{array}
\right]
\end{equation}

\subsection{Controllability}
From Equation (\ref{eq:linearStateSpaceModel}), \(\dot{X}=AX+BU\), the \(A\) and \(B\) matrices can be determined as described the in previous section. Using these \(A\) and \(B\) matrices the controllability test matrix from Equation (\ref{eq:controlTestMatrix}) can be calculated \cite{Friedland1986}.

\begin{equation}
\label{eq:controlTestMatrix}
Q =
\left[
\begin{array}{ccccc}
B & AB & ... & A^{10}B & A^{11}B 
\end{array}
\right]
\end{equation}

It is found that \(\text{rank}\left(Q\right)=12\) meaning that the system is controllable and that designing of the state-space controller can continue unhindered.

\section{\uppercase{System Stabilization - Pole Placement}}
\noindent To stabilize the system with a control law, first the stability of the open loop system must be assessed. It is determined that \(\text{det}\left(sI-A\right)=s^{12}\), which means that there are twelve poles at zero and the system is marginally stable. This is not a desirable behavior, therefore a gain matrix for a closed loop control law must be determined to produce desirable system dynamics and stability.

\subsection{Closed Loop Stabilization}
\noindent The feedback law that was selected to produce desirable system behavior is 
\begin{equation}
\label{eq:controlLaw}
U=-GX
\end{equation}
where  \(G\) is the \(4\times12\) gain matrix. The new closed loop system is thus \(\dot{X}=A_cX\) where \(A_c=\left(A-BG\right)\). To simplify the selection of the control gains it is assumed that: i) vertical control is mainly a function of \(\dot{z}\) and \(z\), ii) roll control is mainly a function of \(\dot{y}\), \(y\), \(\omega_x\). and \(\phi\) (roll), iii) pitch control is mainly a function of \(\dot{x}\), \(x\), \(\omega_y\), and \(\theta\) (pitch), and iv) yaw control is mainly a function of \(\omega_z\) and \(\psi\) (yaw). This leads to a gain matrix of the form seen in Equation (\ref{eq:gainMatrixStructure}).

\begin{equation}
\label{eq:gainMatrixStructure}
G^T=\left[
\begin{array}{cccc}
0 & 0 & g_7 & 0 \\
0 & g_3 & 0 & 0 \\
g_1 & 0 & 0 & 0 \\
0 & 0 & g_8 & 0 \\
0 & g_4 & 0 & 0 \\
g_2 & 0 & 0 & 0 \\
0 & g_5 & 0 & 0 \\
0 & 0 & g_9 & 0 \\
0 & 0 & 0 & g_{11} \\
0 & g_6 & 0 & 0 \\
0 & 0 & g_{10} & 0 \\
0 & 0 & 0 & g_{12} \\
\end{array}
\right]
\end{equation}

The dynamic behavior and the stability of this system is defined by the poles of this system, therefore one would typically think the next step is to select a set of desirable poles for this system. The problem that arises is that while ideally a direct solution for the gains in this matrix can be found from the desired closed loop poles, the characteristic equation found from \(\text{det}\left(sI-A_c\right)\) produces a set of non-linear equations that are difficult to analytically solve. Since all that is needed is a set of poles with good system behavior, an analytical solution is not necessary if a set of poles can be found numerically that produces good system behavior. Therefore, \textit{Simulated Annealing} was used to solve for a set of gains that would match a set of pole criteria.

To create the pole selection criteria it is assumed that a pair of dominant poles will exist that define the majority of the dynamic behavior of the system. This gives a good starting point for selecting a set of poles. In this paper the \(2\%\) setting time of this dominant pole system is defined by \(t_s \approx 4/\zeta\omega_n\) (where \(\zeta\) and \(\omega_n\) are the damping ratio and natural frequency of the second order system, respectively) \cite{Ogata2010}, the \(10\%\) to \(90\%\) rise time is \(t_r \approx 1.8/\omega_n\) \cite{Franklin2010}, and the overshoot is \(OS\% \approx  100\%\times e^{-(\zeta\pi/\sqrt{1-\zeta^2})}\) \cite{Ogata2010}. 

After careful inspection it was observed that a pair of dominant poles with real components located at \(-6\) and low damping ratios of (\(\zeta=0.1\)) have \(t_s \approx 0.67\text{s}\), \(t_r \approx 0.03 \text{s}\), and \(\%OS \approx 73\%\). Similarly, a pair of dominant poles with real components located at \(-6\) and high damping ratios of (\(\zeta = 1.0\)) have \(t_s \approx 0.67\text{s}\), \(t_r \approx 0.3 \text{s}\), and \(\%OS \approx 0\%\). Therefore it was concluded that if the poles of the system were all less than \(-6\) in the real component, that it was likely the overall system would have \(t_s < 1.0\text{s}\), \(t_r < 0.5\text{s}\), and \(\%OS < 100\%\). This was deemed to be sufficient performance. To find the poles it was also determined that the poles should not be too far to the left to prevent excessive control effort, so poles were searched for with real components in the range of \(\ge -6\) and \(\le -30\). Using the Simulated Annealing search algorithm a set of gains for the gain matrix \(G\) were found that satisfied these criteria. Equation (\ref{eq:gainValues}) shows the gain values found, and Equation (\ref{eq:clPoles}) shows the values of the corresponding closed loop poles.

\begin{equation}
\label{eq:gainValues}
\left[
\begin{array}{c}
g_1 \\
g_2 \\
g_3 \\
g_4 \\
g_5 \\
g_6 \\
g_7 \\
g_8 \\
g_9 \\
g_{10} \\
g_{11} \\
g_{12}
\end{array}
\right]
=
\left[
\begin{array}{c}
32.8 \\
608.0 \\
394.5 \\
862.9 \\
47.1 \\
657.9 \\
-397.8 \\
-1124.2 \\
39.4 \\
552.7 \\
30.1 \\
623.4
\end{array}
\right]
\end{equation}

\begin{equation}
\label{eq:clPoles}
\text{Closed Loop Poles}
=
\left[
\begin{array}{c}
-28.32 \\
-20.36 \\
-6.47 \\
-6.28 \\
-16.40 + 18.41i \\
-16.40 - 18.41i \\
-15.05 + 19.92i \\
-15.05 - 19.92i \\
-6.29 + 6.65i \\
-6.29 - 6.65i \\
-6.25 + 2.92i \\
-6.25 - 2.92i
\end{array}
\right]
\end{equation}

At this point it is interesting to notice that this gain matrix has been derived independent of any parameters of the quadrotor. This means that this gain matrix is valid for a wide variety of quadrotor configurations which agree with the assumptions and control structure.

\section{\uppercase{Error Tracking Controller Design}}
\label{sec:errorTrackingDesign}
\noindent With the closed loop system stabilized, it is now desirable to modify the controller to allow the quadrotor to track a reference position. To do this a new closed loop system is designed such that \(\dot{e}=A_c e\) is the new system, where \(e\) is the error state vector for the control system. This results in the new control law as defined in Equation (\ref{eq:errorControlLaw}).

\begin{equation}
\label{eq:errorControlLaw}
U=-Ge
\end{equation}

The next step is to define the error state vector \(e\). For this error tracking system it is desirable that both a set position in inertial space \((x,y,z)\) and the UAVs yaw \((\psi)\) can be tracked. All other quadrotor states are ideally zero (angular and linear speeds are zero). This would seem to lead to an easy conclusion that the error vector should be the same as \(X\), except that \((x,y,z,\psi)\) are subtracted  by their desired reference values. The problem with this error vector design is that with the linear model the control thinks that it can just pitch or roll a large amount to increase the linear speeds to large values. This pushes the pitch and roll outside of the linear range of the controller, causing instability and an inevitable crash of the quadrotor. To account for this a better error vector is used where the linear speeds and angular yaw rate are given errors proportional to the position and yaw errors (Equation (\ref{eq:errorVector}). When setting a position it is more likely that a user will be specifying an inertial position coordinate; however, since the quadrotor works in body frame for \((x,y,z)\) the rotation to body frame must be applied. This is performed using Equation (\ref{eq:posRotation}). In Equation (\ref{eq:errorVector}) the parameters \(k_1\) and \(k_2\) are proportionality constants for the position and yaw errors. While in this paper both \(k_1\) and \(k_2\) were set to \(1.0\), which produced good results, further tuning of these parameters can be performed.

\begin{equation}
\label{eq:errorVector}
e
=
\left[
\begin{array}{c}
\dot{x}+k_1(x-x_{B_{des}}) \\ 
\dot{y}+k_1(y-y_{B_{des}}) \\
\dot{z}+k_1(z-z_{B_{des}}) \\
0 \\
0 \\
0 \\
\omega_x \\
\omega_y \\
\omega_z+k_2(\psi-\psi_{des}) \\
\phi \\
\theta \\
0
\end{array}
\right]
\end{equation}

\begin{equation}
\label{eq:posRotation}
\left[
\begin{array}{c}
x_{B_{des}} \\
y_{B_{des}} \\
z_{B_{des}}
\end{array}
\right]
=
R_{IB}
\left[
\begin{array}{c}
x_{I_{des}} \\
y_{I_{des}} \\
z_{I_{des}}
\end{array}
\right]
\end{equation}

With Equation (\ref{eq:errorVector}) the input vector \(U\) can be determined using the gain matrix \(G\), and consequently the desired rotors speeds (\(\omega_{M_1},\omega_{M_2},\omega_{M_3},\omega_{M_4}\)) can be solved from the values of \(U\).

In addition to this error vector (Equation (\ref{eq:errorVector})), to further improve stability of the system, the maximum linear speed error is saturated at \(1.0\text{m/s}\) and the maximum angular rate error is saturated at \(3.14\text{rad/s}\). These values are subject to further adjustments (depending on the quadrotor properties, especially speed); however, for the purpose of this paper these values produced good results as shown in Section \ref{sec:simulationErrorTracking}.

\section{\uppercase{Simulation}}
\noindent With the stabilized controller and the error tracking controller both defined, the results of simulating the system are now presented. The quadrotor parameters used during simulations are those obtained by \cite{Harandi2010}, and are shown in Table (\ref{tab:quadParams}) with inertial matrix (\(I\)) shown in Equation(\ref{eq:inertialmatrix}).

\begin{table}[h]
\caption{Quadrotor Parameters.}\label{tab:quadParams} \centering
\begin{tabular}{|m{4cm}|c|}
\hline
Half Distance between opposite Rotors (L) & \(0.27\text{m}\) \\\hline
Mass (m) & \(1.4\text{kg}\) \\ \hline
Rotor Speed to Torque Constant (k) & \(11 \times 10^{-6} \text{Ns}^2\) \\ \hline
Rotor Speed to Drag Constant (b) & \(1.1 \times 10^{-6} \text{Nms}^2\) \\ \hline
Max Rotation Speed of Motors (\(\omega_{M_{max}}\)) & \(637.75\text{rad}/\text{s}\)\\ \hline
\end{tabular}
\end{table}

\begin{equation}
\label{eq:inertialmatrix}
\footnotesize
\text{I}
=
\left[
\begin{array}{ccc}
8.1 & 0 & 0 \\
0 & 8.1 & 0 \\
0 & 0 & 14.2 
\end{array}
\right] \times 10^{-3} kg \cdot m^2
\end{equation}

\begin{figure}[h]
\centering
\includegraphics[width=2.5in]{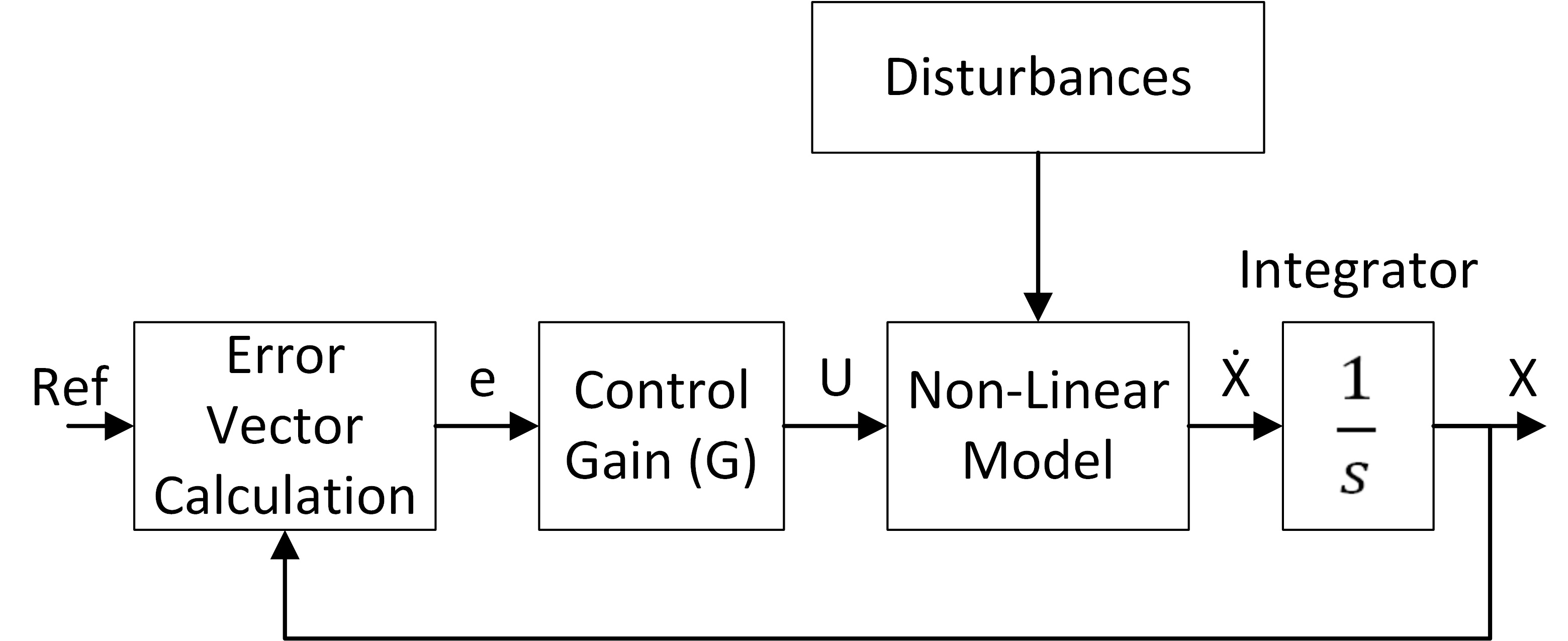}
\caption{Control Structure.}
\label{fig:controlDiagram}
\end{figure}

The model used for the error tracking simulation is shown in Figure \ref{fig:controlDiagram}. The stability analysis diagram is the same except that the "Control Gain" block has a direct input of \(X\). In the simulation the gain matrix, which was designed using linear control design techniques, is used to control the system. The quadrotor is simulated by the non-linear model (Equation (\ref{eq:non-linearModel})). Disturbances are provided to the non-linear model in the stability simulation.

\subsection{Perturbation Response: Stabilized System}
\begin{figure}[h]
\centering
\includegraphics[width=3.25in]{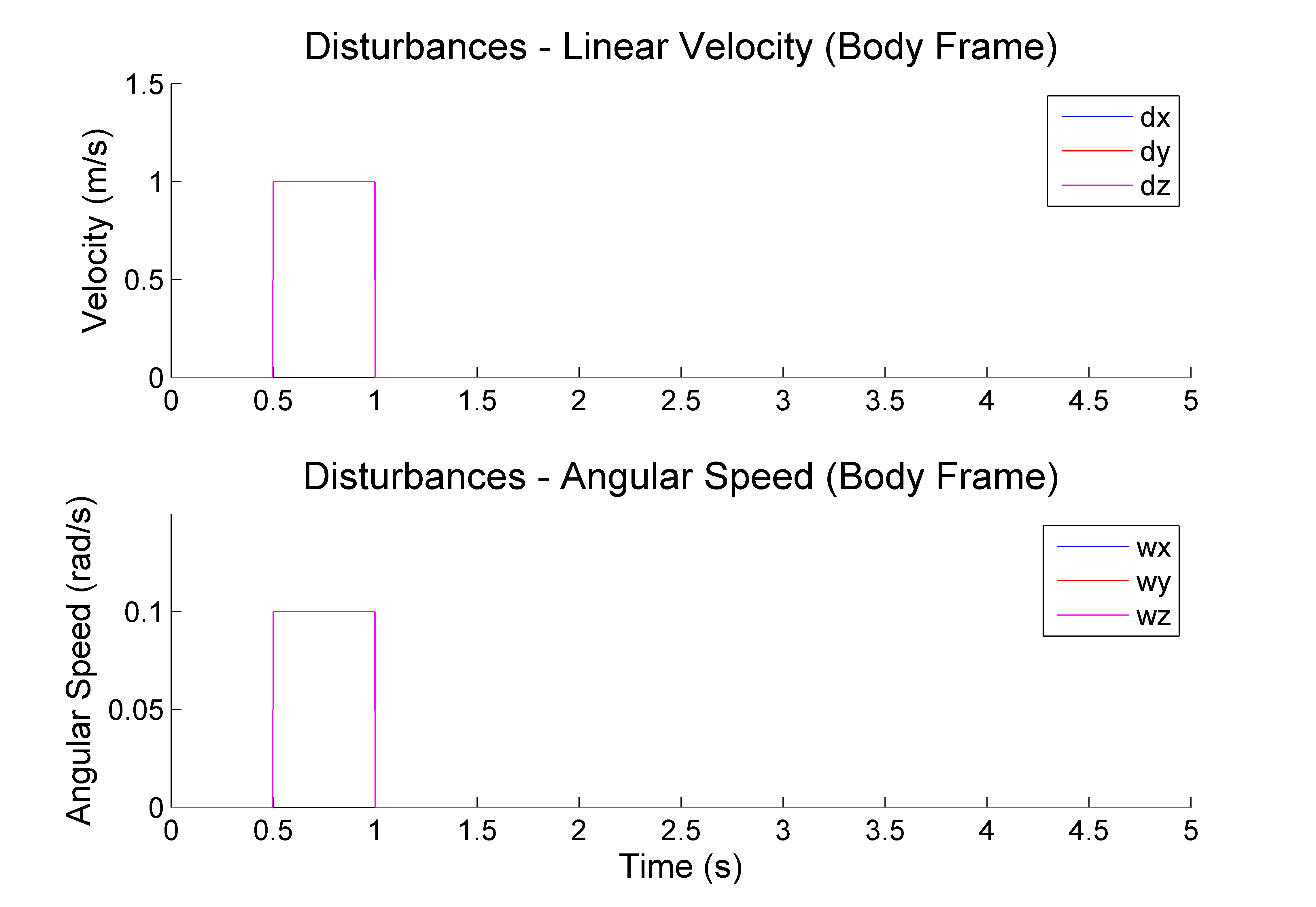}
\caption{Perturbation Simulation: Disturbances.}
\label{fig:perturbDisturb}
\end{figure}

\begin{figure}[h]
\centering
\includegraphics[width=3.25in]{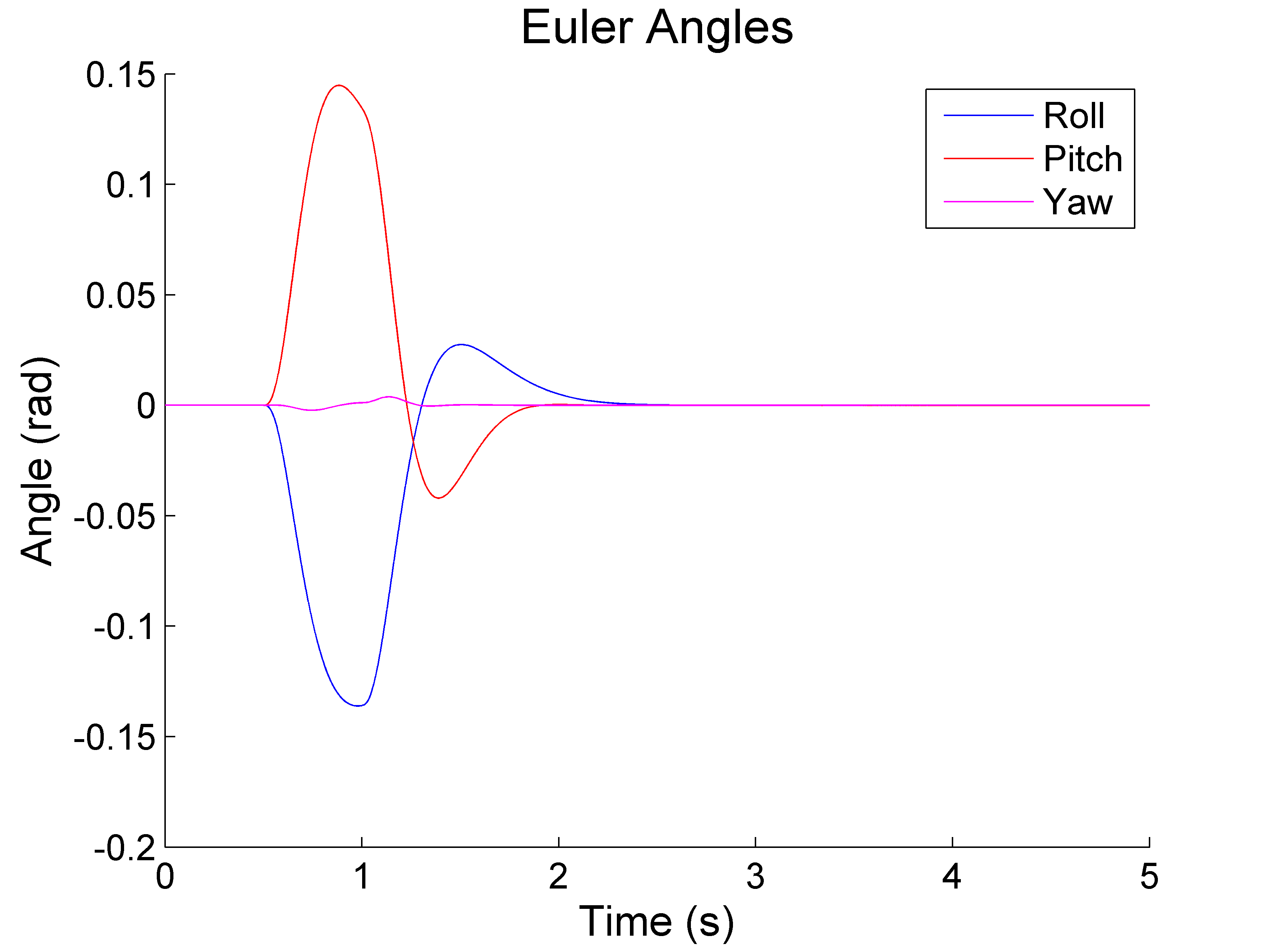}
\caption{Perturbation Simulation: Euler Angles.}
\label{fig:perturbEuler}
\end{figure}

\begin{figure}[h]
\centering
\includegraphics[width=3.25in]{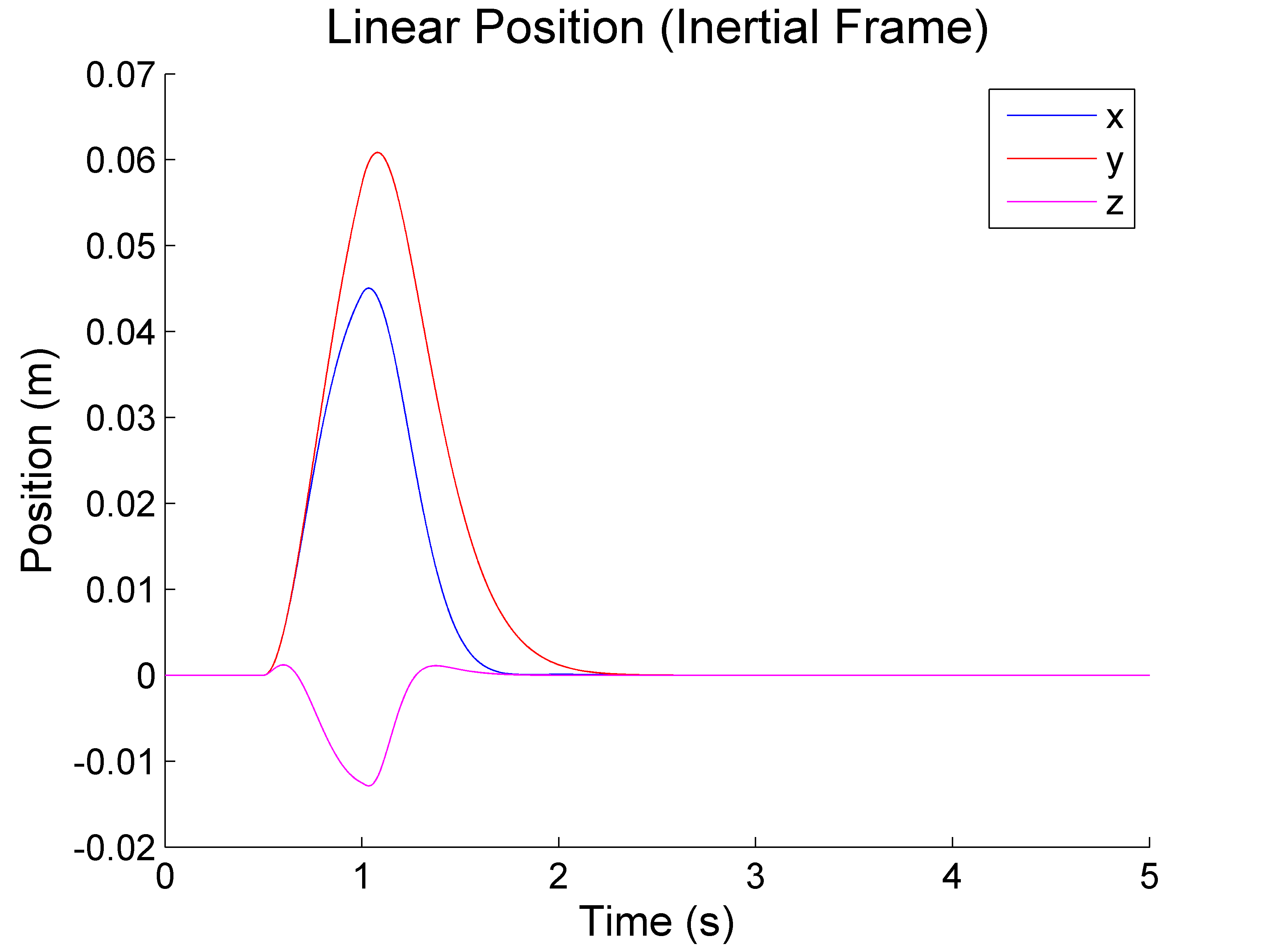}
\caption{Perturbation Simulation: Inertial Positions.}
\label{fig:perturbInertialPosition}
\end{figure}

\noindent To verify the stability of the control system with the selected gains, the closed loop stabilized system (ie. Hover Controller) is subjected to perturbation. The disturbances of \(1\text{m/s}\) for all linear speeds, and \(0.1\text{rad/s}\) for all angular speeds, can be seen in Figure \ref{fig:perturbDisturb}. The system response for the Euler angles can be seen in Figure \ref{fig:perturbEuler}, and for the linear inertial positions in Figure \ref{fig:perturbInertialPosition}.

From Figures \ref{fig:perturbEuler} and \ref{fig:perturbInertialPosition} it can be seen that the system settles quickly after the perturbation ends, reaching steady state in about \(1\text{s}\), which is considered good performance.

\subsection{Random Disturbance Response: Stabilized System}
\begin{figure}[h]
\centering
\includegraphics[width=3.25in]{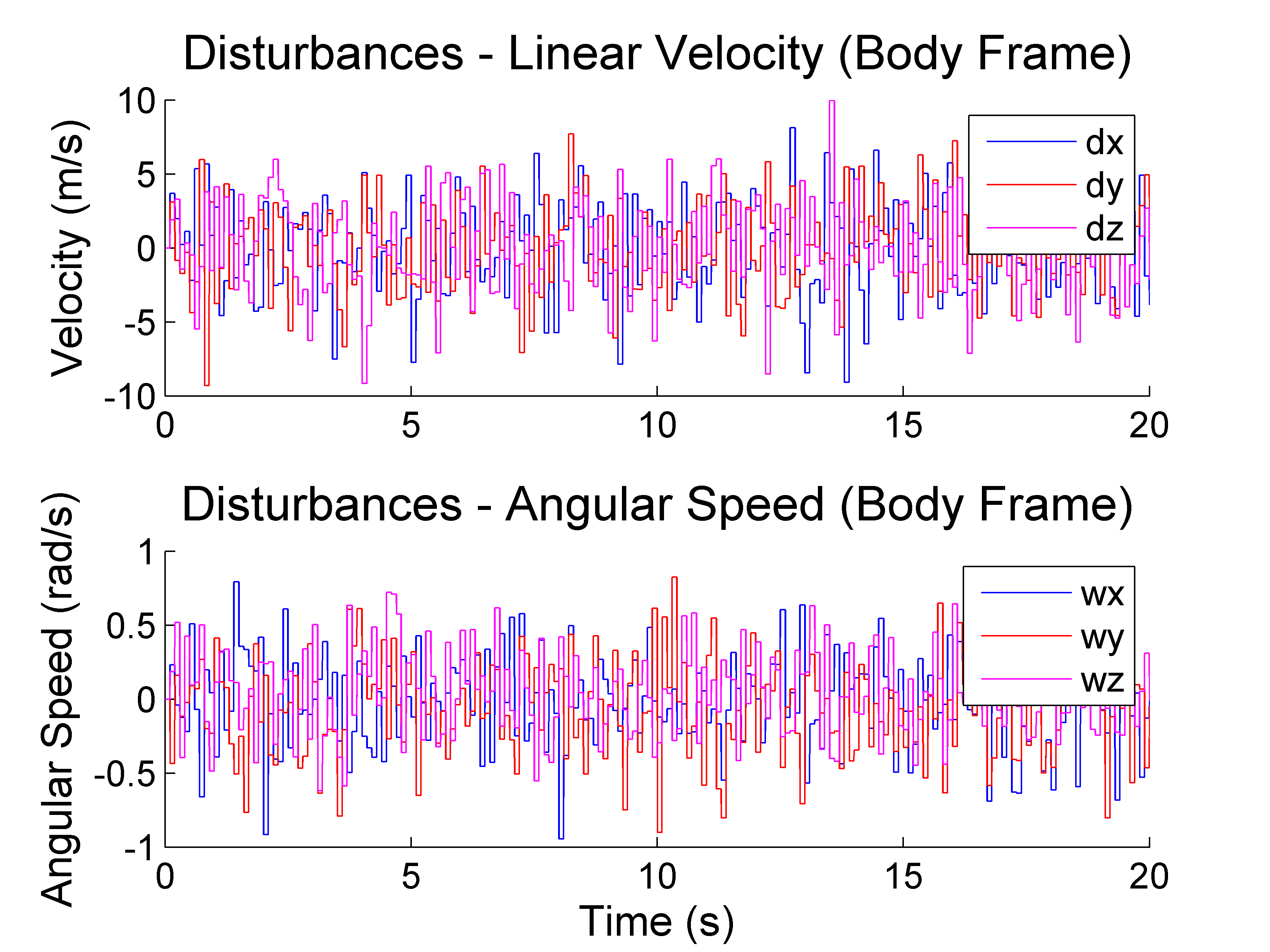}
\caption{Random Simulation: Disturbances.}
\label{fig:randDisturb}
\end{figure}

\begin{figure}[h]
\centering
\includegraphics[width=3.25in]{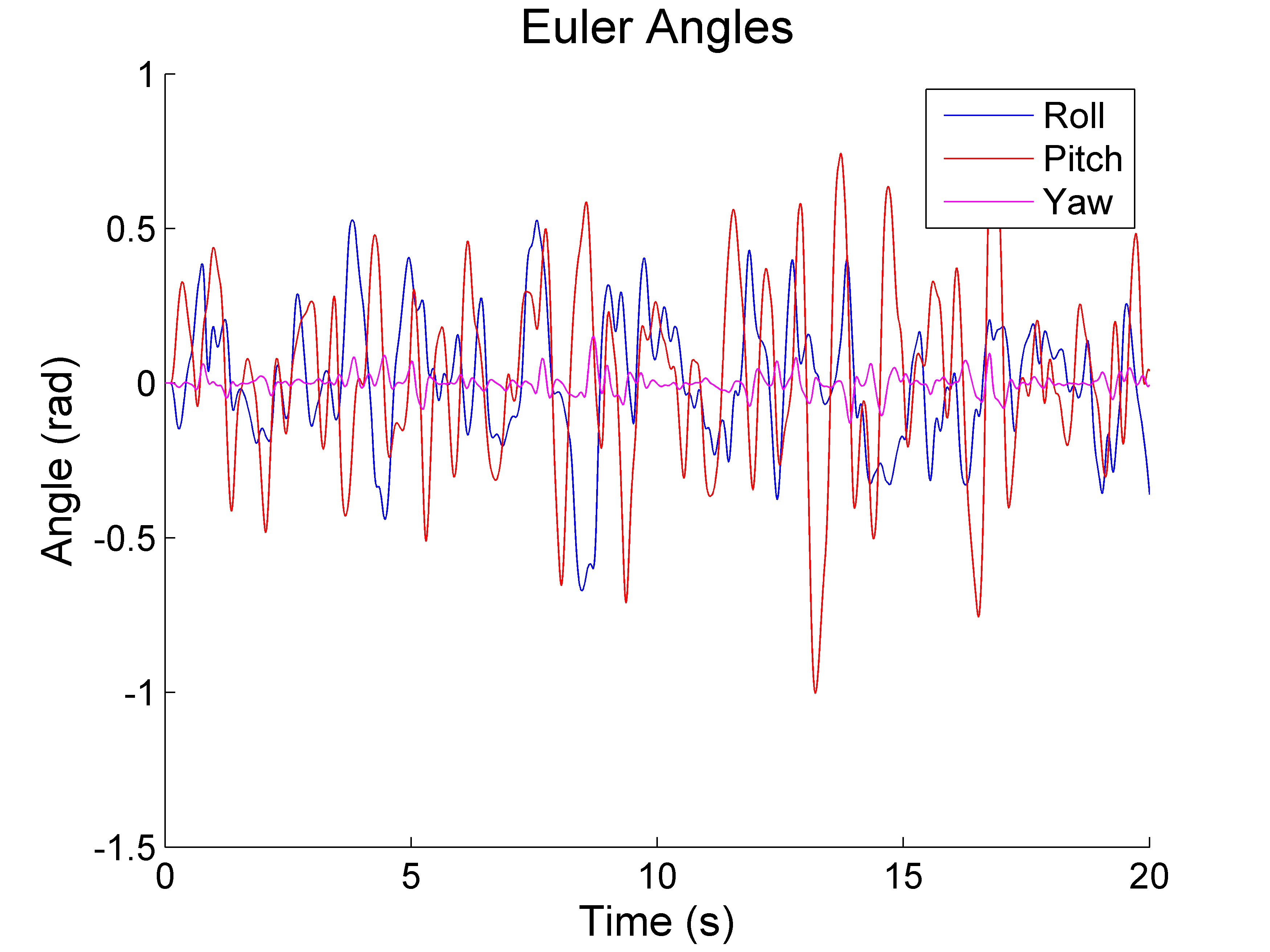}
\caption{Random Simulation: Euler Angles.}
\label{fig:randEuler}
\end{figure}

\begin{figure}[h]
\centering
\includegraphics[width=3.25in]{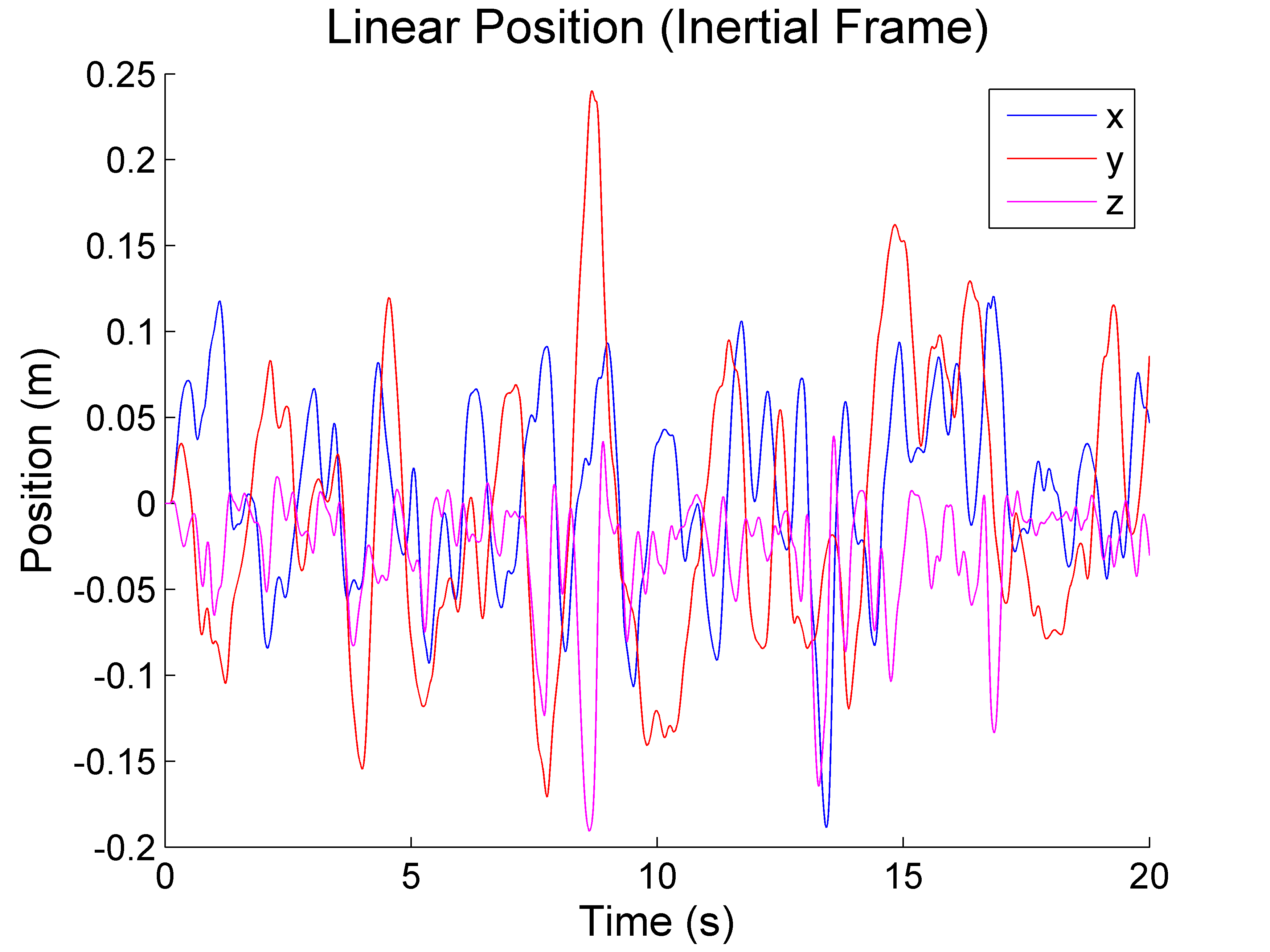}
\caption{Random Simulation: Inertial Positions.}
\label{fig:randInertialPosition}
\end{figure}

\noindent In a second simulation (using Stabilized System, ie. Hover Controller) an extreme case of random disturbances to the linear and angular speeds of the quadrotor (to evaluate the stability in these conditions) was performed. The applied disturbances were normally distributed with a variance of \(10\text{m/s}\) for the linear speed, and \(1\text{rad/s}\) for the angular speed as shown in Figure \ref{fig:randDisturb}.  The system response for the Euler angles can be seen in Figure \ref{fig:randEuler}, and for the linear inertial positions in Figure \ref{fig:randInertialPosition}.

It can be seen from Figures \ref{fig:randEuler} and \ref{fig:randInertialPosition} that while under significant disturbances the system stays stable, and hovers without moving more than \(\pm 0.25\text{m}\) in any direction from its starting point, according to Figure \ref{fig:randInertialPosition}.

\subsection{Error Tracking}
\label{sec:simulationErrorTracking}
\begin{figure}[h]
\centering
\includegraphics[width=3.25in]{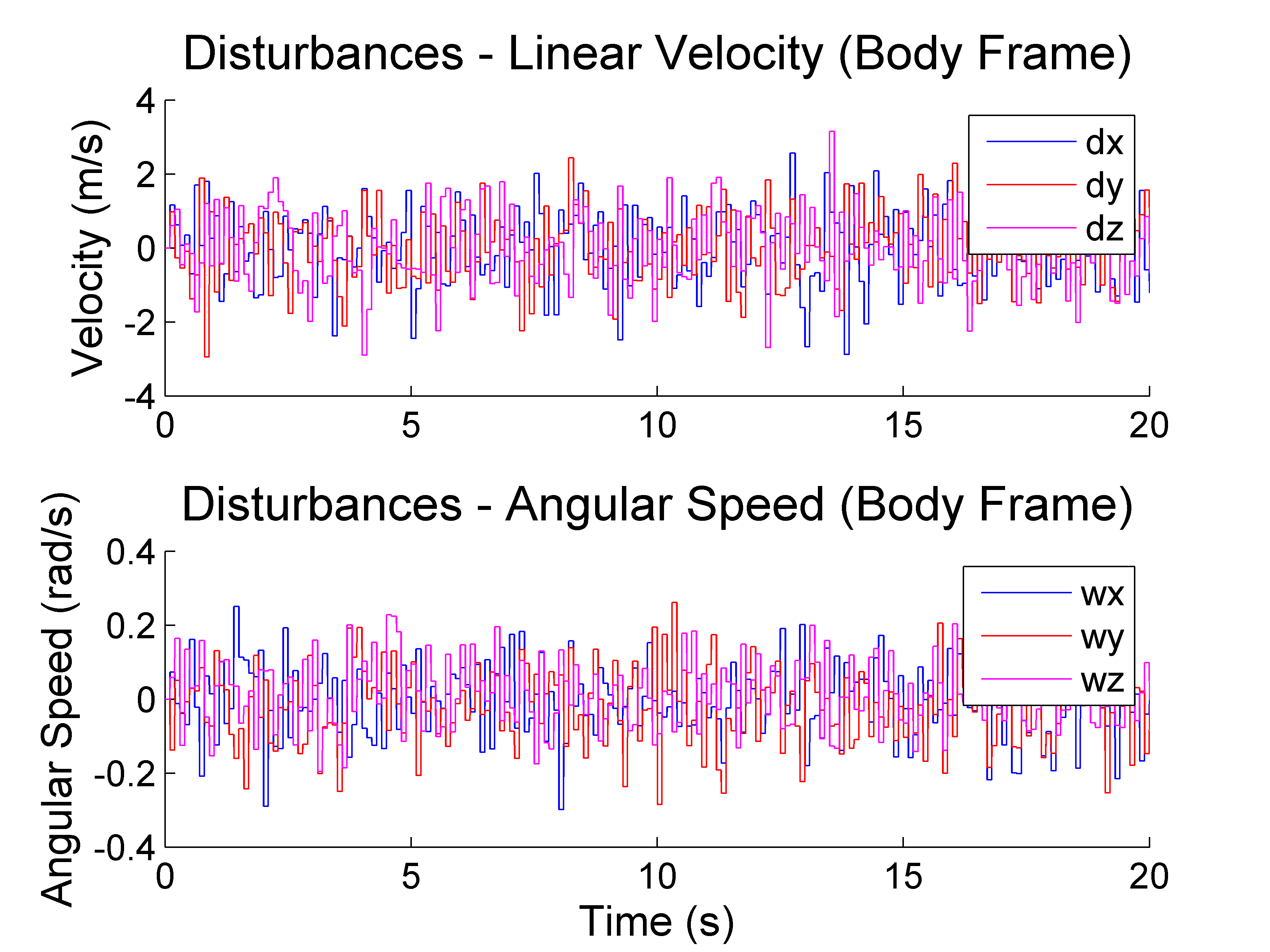}
\caption{Tracking Simulation: Disturbances.}
\label{fig:trackingDisturb}
\end{figure}

\begin{figure}[h]
\centering
\includegraphics[width=3.25in]{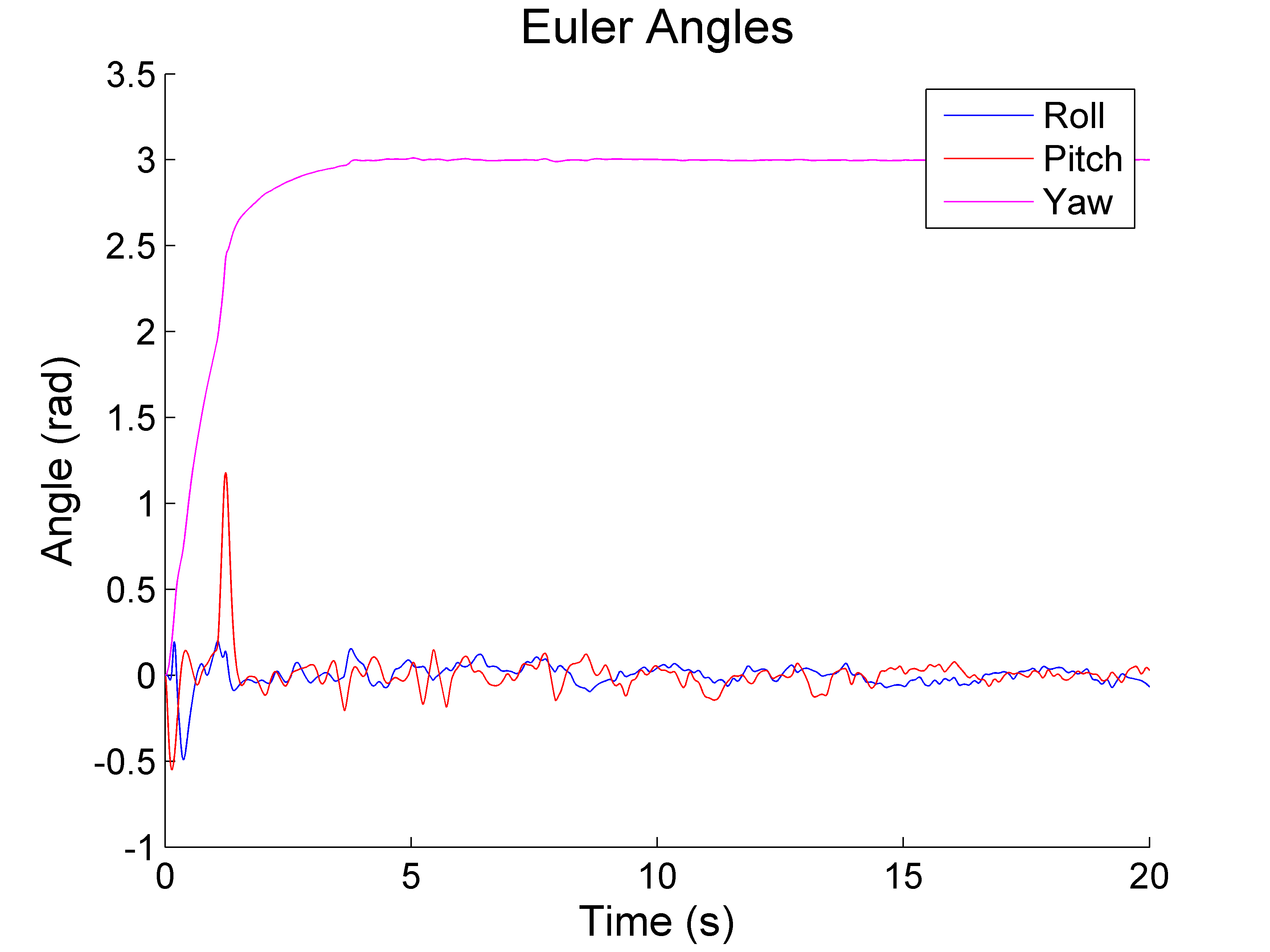}
\caption{Tracking Simulation: Euler Angles.}
\label{fig:trackingEuler}
\end{figure}

\begin{figure}[h]
\centering
\includegraphics[width=3.25in]{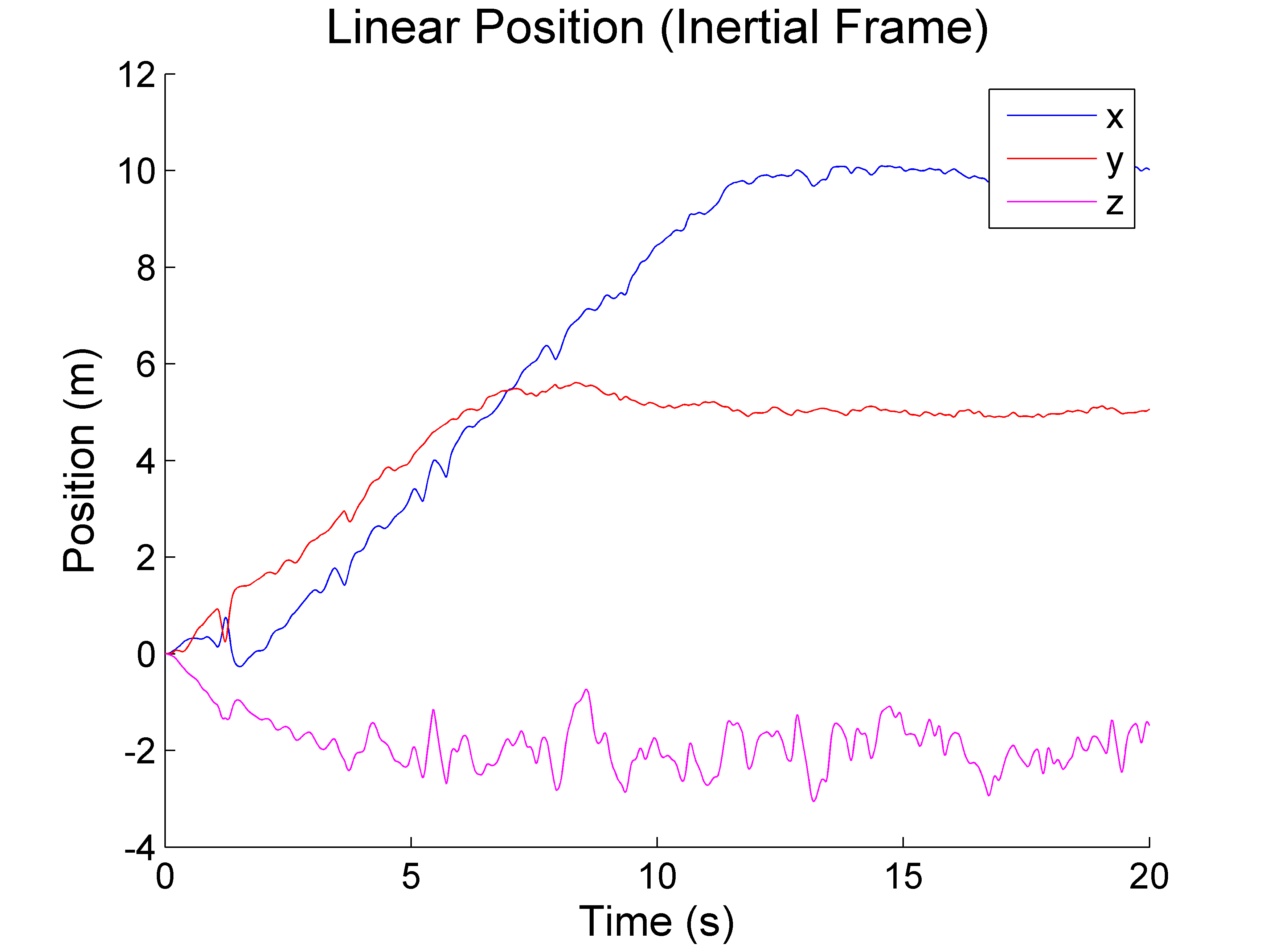}
\caption{Tracking Simulation: Inertial Positions.}
\label{fig:trackingInertialPosition}
\end{figure}

\noindent For the final simulation the error tracking system as designed in Section \ref{sec:errorTrackingDesign} was tested under moderate disturbances to observe the systems ability to track a position and yaw (i.e. desired \((x,y,z)\) position and a desired yaw (\(\psi\))). The disturbances were normally distributed with the linear speeds having a variance of \(1\text{m/s}\), and the angular speeds having a variance of \(0.1\text{rad/s}\) (Figure (\ref{fig:trackingDisturb})). The desired inertial position was set at \((10\text{m},\,5\text{m},\,-2\text{m})\), and the desired yaw was set at \(3\,\text{rad}\). The initial state of the system was at inertial position  \((0\text{m},\,0\text{m},\,0\text{m})\) and yaw \(0\,\text{rad}\). The system response for the Euler angles can be seen in Figure \ref{fig:trackingEuler}, and for the linear inertial positions in Figure \ref{fig:trackingInertialPosition}.

The results of the tracking simulation in Figures \ref{fig:trackingEuler} and \ref{fig:trackingInertialPosition} show that the quadrotor was able track well to the desired linear positions and desired yaw angle. The system stayed stable during this movement and performed well while tracking under moderate disturbances.

\section{\uppercase{Conclusion}}
This paper presented a linear state-space approach at designing a stable hover controller and a stable tracking controller (for inertial position and yaw angle) for a quadrotor Unmanned Aerial Vehicle (UAV). In designing these controllers the gain matrix for the control system was selected using a linearized model of a quadrotor using Simulated Annealing to find gains which would produce a desirable set of closed loop poles. This gain matrix was derived independent of any quadrotor properties (e.g. inertia, dimensions, mass), meaning that it is valid for a wide range of quadrotor configurations provided they agree with the assumptions and control structure. 

The tracking controller designed in this paper works by providing linear velocity and angular velocity error references that are proportional to the linear position and angular position errors as feedback to control system. This type of error feedback prevents problems that would occur with a direct error feedback (i.e. subtracting actual state from reference state) where the system would be pushed out of its operating range and become unstable due to the way the system was linearized for control design.

In simulations the designed gain matrix was used to test system stability under a perturbation disturbance and under random normally distributed disturbances to the linear and angular speeds. Under a perturbation the simulated quadrotor performed well, converging back to zero error in about \(1\text{s}\). The simulated quadrotor also performed well under random normally distributed disturbances (mean of 0, variance of \(10\text{m/s}\) for linear speeds, variance of \(1\text{rad/s}\) for angular speeds) with only \(\pm 0.25\text{m}\) of movement from its zero position in any direction, according to Figure \ref{fig:randInertialPosition}.

The tracking controller was also simulated for a desired position of \((10\text{m},\,5\text{m},\,-2\text{m})\), and the desired yaw of \(3\,\text{rad}\). The simulated quadrotor was subject to moderate normally distributed disturbances with varience of \(1\text{m/s}\) for linear speeds, and a variance of \(0.1\text{rad/s}\) for angular speeds. The tracking controller performed well as the simulated quadrotor achieved the desired position and yaw while staying stable.

The results of this work show that this method of controlling a quadrotor for position tracking and hover stability performs well for this simulation and set of model parameters. This provides support for the hypothesis that since the gain matrix that was derived is independent of the quadrotor properties, it can be applied to any quadrotor system with a similar configuration (given that it agrees with assumptions and control structure). Future work would involve investigation into the effects of state estimation (measurement noise and error) on the performance of this controller, simulation on a wide variety of model parameters, and experimental testing of this control system on a real quadrotor to verify the simulation results.


\end{document}